\documentstyle[preprint,tighten,aps,floats,psfig]{revtex}
                                                                                
\begin{document}
                                                                                
                                                                                
%
%
\preprint{Fermilab-Pub-97/252-E, D0Pub-97-5}
\title{Search for Scalar Leptoquark Pairs Decaying to Electrons
and Jets in $\overline{p}p $\ Collisions}

\author{The D\O\ Collaboration
\footnote{Authors listed on following page. \hfil\break
Submitted to Physical Review Letters.}}
\address{Fermi National Accelerator Laboratory, Batavia, Illinois 60510}
                                                                                
\maketitle
\centerline{\today}
                                                                                
\begin{abstract}
We have searched for the pair production of first generation scalar
leptoquarks in the $eejj$ channel using the full data set (123 pb$^{-1}$)
collected with the D\O\ detector at the Fermilab Tevatron during 1992--1996.
We observe no candidates with an expected background of approximately
0.4 events.  Comparing the experimental 95\% confidence level upper limit
to theoretical calculations of the cross section with the assumption of a 
100\% branching fraction to $eq$, we set a lower limit on the mass of a first
generation scalar leptoquark of 225 GeV/$c^2$.  The results of this analysis
rule out the interpretation of the excess of high $Q^2$ events at HERA
as leptoquarks which decay exclusively to $eq$.
\end{abstract}

%
\newpage
                                                                                
{\small
%
\begin{center}
B.~Abbott,$^{28}$                                                             
M.~Abolins,$^{25}$                                                            
B.S.~Acharya,$^{43}$                                                          
I.~Adam,$^{12}$                                                               
D.L.~Adams,$^{37}$                                                            
M.~Adams,$^{17}$                                                              
S.~Ahn,$^{14}$                                                                
H.~Aihara,$^{22}$                                                             
G.A.~Alves,$^{10}$                                                            
E.~Amidi,$^{29}$
% ...

(D\O\ Collaboration)
\end{center}

\newpage
{\it{
\centerline{$^{1}$Universidad de los Andes, Bogot\'{a}, Colombia}             
\centerline{$^{2}$University of Arizona, Tucson, Arizona 85721}               
\centerline{$^{3}$Boston University, Boston, Massachusetts 02215}             
\centerline{$^{4}$Brookhaven National Laboratory, Upton, New York 11973}      
\centerline{$^{5}$Brown University, Providence, Rhode Island 02912}           
\centerline{$^{6}$Universidad de Buenos Aires, Buenos Aires, Argentina}       
\centerline{$^{7}$University of California, Davis, California 95616}          
\centerline{$^{8}$University of California, Irvine, California 92697}         
\centerline{$^{9}$University of California, Riverside, California 92521}      
\centerline{$^{10}$LAFEX, Centro Brasileiro de Pesquisas F{\'\i}sicas,        
                  Rio de Janeiro, Brazil}                                     
\centerline{$^{11}$CINVESTAV, Mexico City, Mexico}                            
\centerline{$^{12}$Columbia University, New York, New York 10027}             
\centerline{$^{13}$Delhi University, Delhi, India 110007}                     
\centerline{$^{14}$Fermi National Accelerator Laboratory, Batavia,            
                   Illinois 60510}                                            
\centerline{$^{15}$Florida State University, Tallahassee, Florida 32306}      
\centerline{$^{16}$University of Hawaii, Honolulu, Hawaii 96822}              
\centerline{$^{17}$University of Illinois at Chicago, Chicago,                
                   Illinois 60607}                                            
\centerline{$^{18}$Indiana University, Bloomington, Indiana 47405}            
\centerline{$^{19}$Iowa State University, Ames, Iowa 50011}                   
\centerline{$^{20}$Korea University, Seoul, Korea}                            
\centerline{$^{21}$Kyungsung University, Pusan, Korea}                        
\centerline{$^{22}$Lawrence Berkeley National Laboratory and University of    
                   California, Berkeley, California 94720}                    
\centerline{$^{23}$University of Maryland, College Park, Maryland 20742}      
\centerline{$^{24}$University of Michigan, Ann Arbor, Michigan 48109}         
\centerline{$^{25}$Michigan State University, East Lansing, Michigan 48824}   
\centerline{$^{26}$Moscow State University, Moscow, Russia}                   
\centerline{$^{27}$University of Nebraska, Lincoln, Nebraska 68588}           
\centerline{$^{28}$New York University, New York, New York 10003}             
\centerline{$^{29}$Northeastern University, Boston, Massachusetts 02115}      
\centerline{$^{30}$Northern Illinois University, DeKalb, Illinois 60115}      
\centerline{$^{31}$Northwestern University, Evanston, Illinois 60208}         
\centerline{$^{32}$University of Notre Dame, Notre Dame, Indiana 46556}       
\centerline{$^{33}$University of Oklahoma, Norman, Oklahoma 73019}            
\centerline{$^{34}$University of Panjab, Chandigarh 16-00-14, India}          
\centerline{$^{35}$Institute for High Energy Physics, 142-284 Protvino,       
                   Russia}                                                    
\centerline{$^{36}$Purdue University, West Lafayette, Indiana 47907}          
\centerline{$^{37}$Rice University, Houston, Texas 77005}                     
\centerline{$^{38}$Universidade do Estado do Rio de Janeiro, Brazil}          
\centerline{$^{39}$University of Rochester, Rochester, New York 14627}        
\centerline{$^{40}$CEA, DAPNIA/Service de Physique des Particules,            
                   CE-SACLAY, Gif-sur-Yvette, France}                         
\centerline{$^{41}$Seoul National University, Seoul, Korea}                   
\centerline{$^{42}$State University of New York, Stony Brook,                 
                   New York 11794}                                            
\centerline{$^{43}$Tata Institute of Fundamental Research,                    
                   Colaba, Mumbai 400005, India}                              
\centerline{$^{44}$University of Texas, Arlington, Texas 76019}               
\centerline{$^{45}$Texas A\&M University, College Station, Texas 77843}       

}}

                                                                                
\newpage
Leptoquarks (LQ) are hypothesized exotic color-triplet bosons which couple to
both quarks and leptons.  They appear in extended gauge theories and composite
models \cite{generic_lq} and  have attributes of both quarks and leptons
such as color,  fractional electric charge, and lepton and baryon quantum
numbers.
Leptoquarks with universal couplings to all flavors would give rise to
flavor-changing neutral currents and are severely constrained by studies of
low energy phenomena \cite{low_e}.
Therefore, only leptoquarks which couple
within a single generation are considered here.  The H1 and ZEUS
experiments at
HERA have reported an excess of events at high $Q^2$ in $e^+p$
collisions \cite{hera}.  One possible interpretation of these events is
resonant production of first generation leptoquarks
\cite{hewett} at a mass near 200 GeV/$c^2$.
To date, no excess has been observed in $e^-p$ collisions \cite{zeus_lq}.
A straightforward leptoquark explanation then requires the leptoquarks
to decay to $eq$ with a branching fraction of 100\% \cite{hewett}.
                                                                                
The CDF and D\O\ collaborations have published the results of searches
for first generation leptoquarks in $\overline{p}p $\ collisions
using data collected prior to 1994
\cite{old_lq}.  The H1 and ZEUS experiments at HERA have published lower limits
on the mass of a first generation leptoquark which depend on the unknown
LQ-$e$-$q$\ coupling \cite{zeus_lq,hera_lq}.  Experiments at LEP have searched
for leptoquarks in decays of the $Z$ boson \cite{lep_lq}.
                                                                                
This Letter describes a search for the pair production of first generation
scalar leptoquarks in the $eejj + X$ final state
using $123 \pm 7$ pb$^{-1}$\ of data collected at the Fermilab Tevatron with
$\sqrt{s} = 1.8$ TeV during 1992--1996.
The D\O\ detector and data acquisition system are described in detail in
Ref.~\cite{d0nim}.  The detector consisted of three major subsystems:
a uranium/liquid-argon calorimeter; central tracking detectors, including
a transition radiation detector; and a muon spectrometer.
                                                                                
A base data sample of 101 events with two electrons and two or more jets was
selected.  Electrons were identified
by their longitudinal and transverse shower profiles in the calorimeter
and by the fraction of their energy deposited in its electromagnetic
section.  The electrons
were required to be isolated from other energy depositions
and to have pseudorapidity  $|{\eta}| < 1.1$ or
$1.5 < |{\eta}| < 2.5$.
In addition, at
least one electron was required to have a matching track in the central
tracking detectors and to satisfy ionization requirements in
the tracking chambers and transition radiation detector.
Jets were reconstructed using a cone algorithm of radius
${\cal{R}} \equiv \sqrt{(\Delta\phi)^2 +(\Delta\eta)^2} = 0.7$, where
$\phi$ is the azimuthal angle, and required to have
$|{\eta}| < 2.5$.  The electrons were required to be separated from jets
by ${\cal{R}} > 0.7$.  Electrons were required to have transverse
energy $E_T^e > 20$~GeV and jets to have $E_T^j > 15$~GeV.
The kinematic quantities were calculated using the vertex determined by
the electrons.  Events whose
$ee$ invariant mass lies between $82 < M_{ee} < 100$ GeV/$c^2$\
($Z$ boson region) were rejected.
The efficiency of the electromagnetic trigger used to collect the base
data sample exceeded 99\% for the leptoquark mass range addressed by this
analysis.
                                                                                
Monte Carlo (MC) signal samples were generated for leptoquark masses between
120 and
260 GeV/$c^2$\ using the {\footnotesize{ISAJET}} \cite{isajet} event generator
and a detector simulation based on the
{\footnotesize{GEANT}} \cite{geant} program.
Leptoquark production cross sections were taken from the recently available
next-to-leading order (NLO) calculations of Ref.~\cite{kraemer}.
The primary backgrounds to the $eejj$ decay mode are
Drell-Yan + 2 jets production (DY),
$t\overline{t}$\ production, and misidentified multijet events.  Monte Carlo
samples for the DY events were generated using {\footnotesize{ISAJET}}.
The DY cross section normalization was fixed by comparing the MC events with
$Z + 2$ jets data in the $Z$ boson region.
Top quark events were generated using the {\footnotesize{HERWIG}}
\cite{herwig} program at a top quark mass of 170 GeV/$c^2$\ and all dilepton
final states were included.  The D\O\ measured $t\overline{t}$\ production
cross section
of $5.5 \pm 1.8$ pb at a top quark mass of 173.3 GeV/$c^2$\ was used
\cite{top_sigma}.
The multijet background
was estimated from a sample of events with four or more jets in which the
probability
for two jets or photons to be misidentified as electrons was weighted
by the number of
jets in the event which passed the electron $E_T$\ and $\eta$ requirements.
This misidentification
probability was calculated from a sample of events with three jets to be
$(3.50 \pm 0.45)\times 10^{-4}$ for an electron with a reconstructed
track and $(1.25 \pm 0.13)\times 10^{-3}$ for an electron without
a reconstructed track.  The errors on these probabilities reflect a slight
dependence on the
jet $E_T$ and $\eta$. The signal and
background samples were required to pass the kinematic selection criteria
that defined the base sample.
                                                                                
To search for leptoquarks, a random grid search method  \cite{rgsearch}
was used to optimize cuts on the data and MC samples.
Consistent results were obtained using a neural network \cite{nn}.
Two criteria were used to optimize event selection, one designed for
discovery and the other for limit setting.
Once it became clear that the data
did not contain evidence for leptoquark production, the limit setting
criterion of a maximum number of signal events
for a fixed number of background events was adopted. The background
level chosen was 0.4 events, corresponding to a 67\% probability that
no such events would be observed.
                                                                                
The set of cuts which optimally separates signal from background
was determined using a systematic search over a grid of possible cuts
with the choice of grid points determined by the distributions of signal
MC events.  Many sets of selection criteria were explored including
combinations of kinematic
quantities and mass-related variables, such as transverse energy and the two
$ej$ invariant masses.
A cut on a single, relatively simple variable,
$S_T \equiv H_T^e + H_T^j$, where
$H_T^e \equiv E_T^{e1} + E_T^{e2}$ and
$H_T^j \equiv \sum_{\text {jets}} E_T^j$, satisified the limit setting
criterion.
Approximately 0.4 background events are expected for $S_T > 350$ GeV.
No events remain in the base data sample after this $S_T$
cut is applied.  Figure~\ref{fig:st} shows the $S_T$ distribution
for the base data sample, the predicted background, and a MC sample of
200 GeV/$c^2$\ LQ events.
The highest value of $S_T$\ seen in the data is 312 GeV.
                                                                                
\begin{figure}\vbox{
\centerline{
\psfig{figure=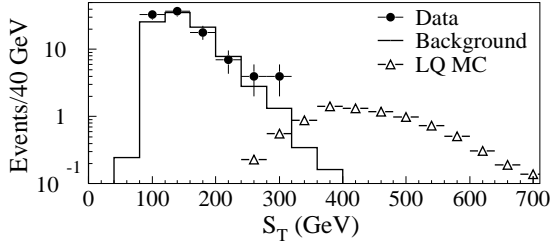,width=3.in}}
\caption{$S_T$ distributions
for background (solid line histogram),
data (solid circles), and
200 GeV/c$^2$\ leptoquark MC events (open triangles).}
\label{fig:st}
}
\end{figure}

For the neural network (NN) analysis, a three layer feed-forward network 
was constructed with two inputs, $H_T^e$ and $H_T^j$, and one output, the 
NN discriminant $D_{N\!N}$.
Figure~\ref{fig:opt}(a) shows the expected distribution in
$H_T^e$ {\sl vs.} $H_T^j$  for the
200 GeV/$c^2$\ MC signal sample; Figs.~\ref{fig:opt}(b) and (c)
show the same distributions for the predicted background and the
base data sample.
The network was trained using the 200 GeV/$c^2$\ LQ sample and the background
samples described above. $D_{N\!N}$ has a range between
0 (background) and 1 (signal).
Figures \ref{fig:opt}(a--c) show
contours corresponding to three values of $D_{N\!N}$.
A background of $\approx0.4$ events is obtained by requiring
$D_{N\!N}>0.95$.  After application of this cut, no events
remain in the data.  As the NN and $S_T$ analyses provide nearly identical
sensitivity, only the simpler $S_T$ analysis
was used for the cross section limit described in this Letter.
                                                                                
\begin{figure}\vbox{
\centerline{
\psfig{figure=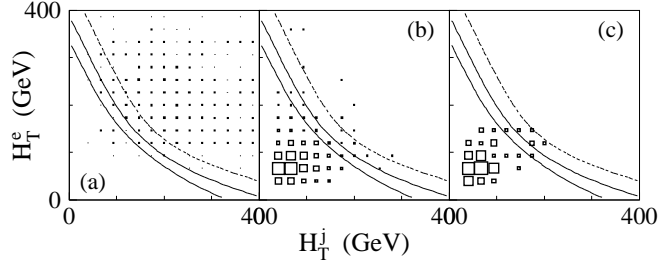,width=3.5in}}
\caption{
$H_T^e$ {\sl vs.} $H_T^j$ for (a) 200 GeV/$c^2$\ LQ events,
(b) predicted background, and (c) base data sample.
The curved lines correspond to $D_{N\!N} = 0.5$, 0.8, and 0.95
(from left to right).  The area of a box is proportional to the number of
events in the bin, with the total number of events normalized
to 123 pb$^{-1}$.}
\label{fig:opt}
}
\end{figure}

The background was estimated for $S_T > 350$\ GeV and is given in
Table \ref{tab:bkg} for the three sources.  The total estimated background
is $0.44 \pm 0.06$ events where the error includes both statistical and
systematic uncertainties.
Included in the systematic error
are the uncertainties in the jet energy scale,
particle identification efficiency, $t\overline{t}$\
production cross section, and luminosity, and the effects
of the choice of parton distribution function
and renormalization and factorization scale $\mu$, gluon radiation, and
MC statistics.

\begin{table}
\caption{Background contributions from individual sources.}
\begin{tabular} { c c }
Background Source           & Number of Events \\ \hline
DY                          & $ 0.18 \pm 0.04 $ \\
$t\overline{t}$             & $ 0.11 \pm 0.04 $ \\
Multijet Misidentification  & $ 0.16 \pm 0.02  $ \\ \hline
Total                       & $ 0.44 \pm 0.06 $ \\
\end{tabular}
\label{tab:bkg}
\end{table}

Modeling of the $S_T$ distribution for high mass
DY events was checked by studying $H_T^e$ and $H_T^j$ separately using
data and MC events in the $Z$ boson mass region.
The average value of $H_T^e$ for high mass DY events (which provide most
of the DY background)
is approximately
250 GeV, corresponding to an $H_T^j$ of approximately 100 GeV for
$S_T = 350$ GeV.  The distribution of $H_T^j$ for high mass DY events is
expected to be similar to that of $Z + 2$\ jets events.  Figure~\ref{fig:ht_z}
shows the $H_T^j$ distributions for $Z + 2$\ jets MC and data.  In the
region corresponding to the $S_T$ cut for high mass DY events
($H_T^j \approx 100$ GeV), the agreement is good.

\begin{figure}\vbox{
\centerline{\psfig{figure=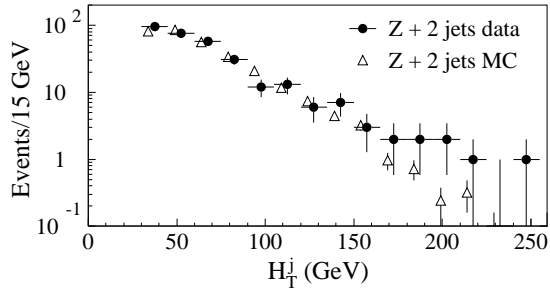,width=3.in}}
\caption{The $H_T^j$ distribution for $Z + 2$ jets data (solid circles)
and MC (open triangles) in the $Z$ boson mass region.  For high mass DY
events, $S_T = 350$ GeV corresponds to $H_T^j \approx 100$~GeV.}
\label{fig:ht_z}
}
\end{figure}
                                                                                
To investigate the background further, constrained mass fits were
performed on the events in the base data sample, on background samples,
and on the 200 GeV/$c^2$\ leptoquark signal sample.
The 3C mass fit was based on the {\footnotesize{SQUAW}}
\cite{squaw} kinematic mass fitting program and required the two $ej$ masses
to be identical.  Use of the fitting
program improves the mass resolution by approximately 10\% over a
simple calculation of the $ej$ invariant masses.
Figures~\ref{fig:mass_dist}(a--c) show $S_T$ as a function of the
fit mass for the estimated background,
200 GeV/$c^2$\ leptoquark events, and the base data sample.
The distribution from the data agrees with that of the
background.  The two highest mass events have low values of $S_T$ and so are
unlikely to be leptoquark events.
Figure~\ref{fig:mass_dist}(d) shows the one dimensional
mass distributions for the same samples.
Inset in Fig.~\ref{fig:mass_dist}(d) are the distributions after a cut
on $S_T > 250$\ GeV.
As can be seen, the data are consistent with the background prediction.
                                                                                
\begin{figure}\vbox{
\centerline{
\psfig{figure=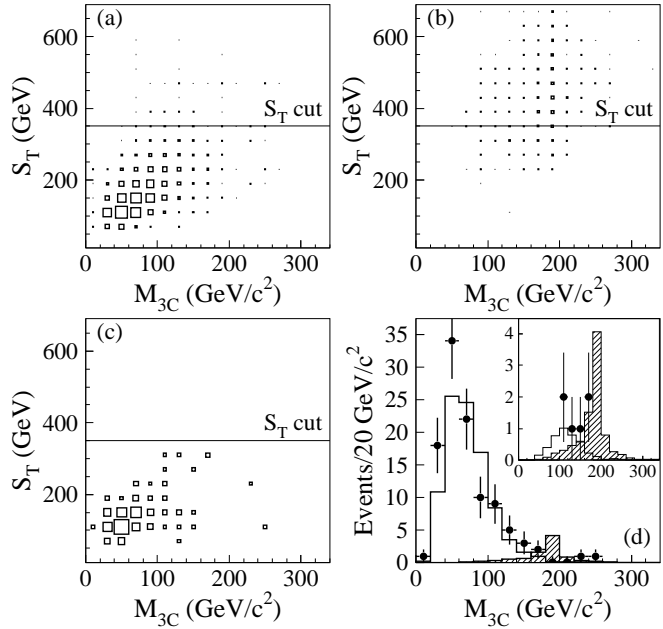,width=3.5in}}
\caption{$S_T$ {\sl vs.}~3C fit mass distributions for (a) background,
(b) 200 GeV/$c^2$\ leptoquarks, and (c) the base data sample. The area
of a box is proportional to the number of events in the bin.
(d) Mass distribution of the events in the base data sample
(solid circles), expected background (solid line histogram), and
200 GeV/$c^2$\ leptoquarks (hatched histogram).
The inset plot shows these distributions for events with $S_T > 250$ GeV.
}
\label{fig:mass_dist}
}
\end{figure}
                                                                                
The dielectron identification efficiency
was determined to be ($73 \pm 4$)\%
using a sample of $Z \to ee + 2$ jets events.
The overall signal detection efficiency is 9--37\%
for leptoquark masses of 120--250 GeV/$c^2$\ (Table II).
We set a 95\% confidence level (CL) upper limit on the cross section
$\sigma$ using
a Bayesian approach
with a flat prior distribution for the signal cross section.
The statistical and systematic uncertainties on the efficiency, the
integrated luminosity, and the
background estimation were included in the limit calculation with
Gaussian prior distributions.  The resulting upper limit on the cross section
is shown in Fig.~\ref{fig:limit} together with the NLO calculation of
Ref.~\cite{kraemer}, and the results are listed
in Table \ref{tab:sigma_details}.
The intersection of our limit curve with the lower edge
of the theory band ($\mu = 2M_{\text{LQ}}$) is at $\sigma = 0.068$ pb,
leading to a lower limit on the mass
of a first generation scalar leptoquark of 225 GeV/$c^2$.
For a branching fraction of 100\% to $eq$, this is our lower limit;
inclusion of additional channels will provide
increased sensitivity in the case where leptoquarks also decay to $\nu q$.
                                                                                
In conclusion, we have excluded the interpretation of the HERA high $Q^2$
events as first generation scalar leptoquarks which decay exclusively to $eq$,
as expected 
in chiral models with no extra fermions or intergenerational mixing.
Using the NLO cross section calculation of Ref.~\cite{kraemer} with
$\mu = 2M_{\text{LQ}}$, and
assuming a 100\% branching fraction to $eq$, the 95\% CL lower limit on
the mass of a first generation scalar leptoquark is 225 GeV/$c^2$.

\begin{table}
\caption{The signal detection efficiency,
the 95\% CL upper limit on the production
cross section, and the $\mu=2M_{\text{LQ}}$ NLO cross sections
from Ref.~\protect\cite{kraemer}. }
\begin{tabular} { c c c c }
Leptoquark  & Signal  &
95\% CL  & NLO Theory  \\
Mass & Efficiency & Upper Limit & Cross Section \\
(GeV/$c^2$) & (\%) & (pb) & (pb) \\
\hline
120 & $ 8.7 \pm 1.4$ & 0.170 & 3.8 \\
160 & $20.8 \pm 3.0$ & 0.113 & 0.68 \\
200 & $31.2 \pm 3.8$ & 0.078 & 0.16 \\
225 & $35.7 \pm 4.3$ & 0.068 & 0.068 \\
250 & $37.2 \pm 4.5$ & 0.066 & 0.030 \\
\end{tabular}
\label{tab:sigma_details}
\end{table}
                                                                                
We thank M.~Kr\"amer for discussions and detailed cross section
information and J.L.~Hewett and T.G.~Rizzo for many useful discussions.
%
We thank the staffs at Fermilab and collaborating institutions for their
contributions to this work, and acknowledge support from the 
Department of Energy and National Science Foundation (U.S.A.),  
Commissariat  \` a L'Energie Atomique (France), 
State Committee for Science and Technology and Ministry for Atomic 
   Energy (Russia),
CNPq (Brazil),
Departments of Atomic Energy and Science and Education (India),
Colciencias (Colombia),
CONACyT (Mexico),
Ministry of Education and KOSEF (Korea),
and CONICET and UBACyT (Argentina).

\begin{figure}\vbox{
\centerline{
\psfig{figure=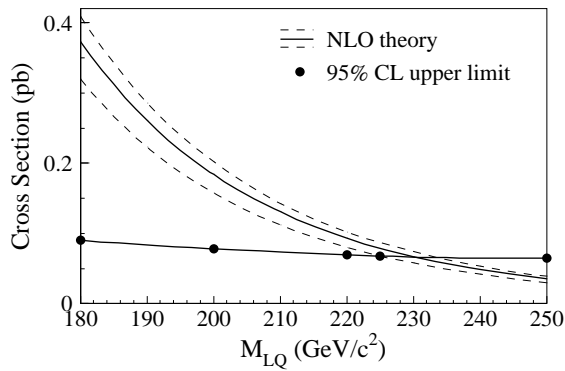,width=3.in}}
\caption{ Upper limit on the leptoquark pair production cross section
for 100\% decay to $eq$. Also shown is the NLO calculation
of Ref.~\protect\cite{kraemer} where the central solid line corresponds to
$\mu = M_{\text{LQ}}$, and the lower and upper dashed lines to
$\mu = 2M_{\text{LQ}}$ and $\mu = M_{\text{LQ}}/2$, respectively. }
\label{fig:limit}
}
\end{figure}

\end{document}